\documentclass[preprint]{elsarticle_new}
\usepackage{amsmath}
\usepackage{amssymb}
\usepackage[final]{epsfig}
\usepackage{color}
\usepackage{graphicx}
\usepackage{lineno}
\newcommand{\beq}{\begin{equation}}
\newcommand{\eeq}{\end{equation}}
\newcommand{\bdi}{\begin{displaymath}}
\newcommand{\edi}{\end{displaymath}}

\newcommand{\bea}{\begin{eqnarray}}
\newcommand{\eea}{\end{eqnarray}}

\newcommand{\ov}{\overline}

\begin{document}

\begin{frontmatter}

\title{Gain fluctuations in a uniform avalanche diode}

\author{W. Riegler}

\address{CERN}

\begin{abstract}
 
 \noindent
We derive the explicit probability distribution of the total avalanche size produced by $m$ primary electron-hole pairs created at an arbitrary position in a uniform avalanche diode. We also derive an operator that generates all raw moments.

\end{abstract}

\end{frontmatter}

\section{Introduction}

\noindent
Gain fluctuations in avalanche diodes were studied in numerous publications \cite{McIntyre1966,Personick1971NewResults,Personick1971GeneralClass,McIntyre1972,Balaban1976,VanVliet1979Part1,VanVliet1979Part2,Chakraborti1984,WindischhoferRiegler2021,WindischhoferRiegler2021_2}. Notably, McIntyre \cite{McIntyre1972} derived the explicit gain distribution for the case where $m$ primary electrons are injected at the boundary of the diode.In this publication, we derive the distribution of the total
avalanche size for one or more primary electron--hole pairs
created at an arbitrary position $x$ in the diode. We also derive an operator that generates all raw moments. The calculation is based on the ionization probability of $\alpha dx$ for electrons and $\beta dx$ for holes. Clearly the assumption of a uniform electric field is idealized and it is well established that for detailed calculations one has to take into account the fact that after an ionizing collision there is a certain distance (dead space) that the electrons and holes have to travel before having sufficient energy to ionize. Dead-space and nonlocal transport effects can be treated using nonlocal ionization models or Monte Carlo simulations.  It is still useful to have analytic expressions for this idealized case because they clearly set the scale of the effects.
\section{Electron-hole avalanche}

\begin{figure}[ht]
 \begin{center}
  \includegraphics[height=6cm]{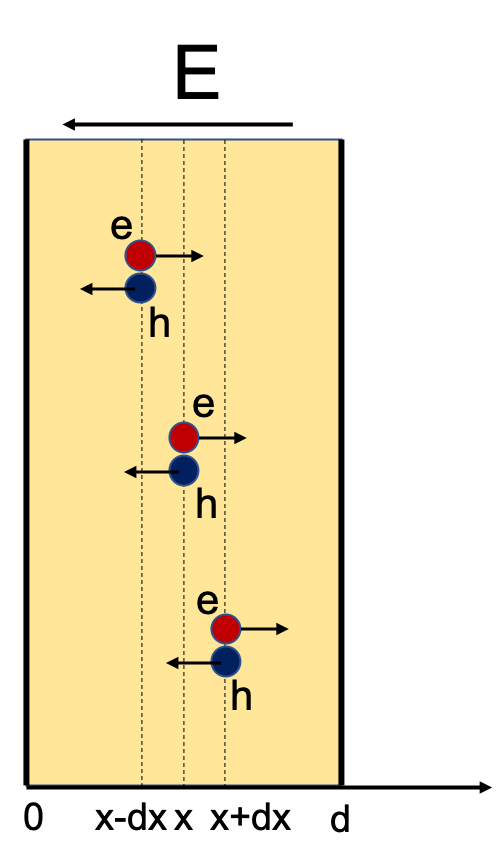}
  \caption{A silicon diode of thickness $d$ with a uniform electric field. A single e-h pair is placed at  position $x$, which starts an avalanche. To establish the equations that define the probability distributions for the gain we relate the gain probabilities for a primary e-h pair at position $x$ to the probabilities at $x-dx$ and $x+dx$.  }
  \label{diode_sketch}
  \end{center}
\end{figure}

\noindent
We assume a diode extending from $x{=}0$ to $x{=}d$ (Fig. \ref{diode_sketch}) and we are interested in the probability $p(n, x)$ of finding a total of $n$ additional electron-hole pairs for an avalanche that has started with a single e-h pair at position $x$. For the calculation we need two additional probabilities, namely the probability $p_e(n,x)$ of finding $n$ additional e-h pairs for an avalanche that starts with a single electron at position $x$ and the probability $p_h(n,x)$ to find $n$ additional e-h pairs for an avalanche that starts with a single hole at position $x$. In the following we define the equations that determine $p(n, x), p_e(n, x)$ and $p_h(n, x)$.
\\ \\
Starting with a single electron at position $x{-}dx$ there are two possibilities to get to $n$ additional e-h pairs. Either the electron moves over a distance $dx$ without multiplication and the electron creates $n$ e-h pairs starting from $x$, or the electron multiplies, we have two electrons and one hole at position $x$, and together they produce $n{-}1$ e-h pairs (one is already produced). This can be expressed as follows:
\beq
       p_e(n, x-dx) = (1-\alpha dx) p_e(n, x) + \alpha dx \sum_{i=0}^{n-1} \sum_{j=0}^{n-1-i} p_e(n-1-i-j, x)p_e(i,  x) p_h(j,  x) 
\eeq
The same reasoning applies for a single hole starting at position $x+dx$
\beq
       p_h(n, x+dx) = (1-\beta dx) p_h(n, x) + \beta dx \sum_{i=0}^{n-1} \sum_{j=0}^{ n-1-i} p_e(n-1-i-j, x)p_h(i,  x) p_h(j,  x) 
\eeq
Expanding for small dx we have
\beq
       -\frac{d p_e(n, x)}{dx}  =  -\alpha  p_e(n, x) + \alpha  \sum_{i=0}^{n-1} \sum_{j=0}^{ n-1- i} p_e(n-1-i-j, x)p_e(i,  x) p_h(j,  x) 
\eeq
\beq
       \frac{d p_h(n, x)}{d x}  =-\beta  p_h(n, x) + \beta  \sum_{i=0}^{n-1} \sum_{j=0}^{n-1- i} p_e(n-1-i-j, x)p_h(i,  x) p_h(j,  x) 
\eeq
The generating function of $p_e(n, x)$ and $p_h(n, x)$ is defined as 
\beq
     P_e(s, x) = \sum_{n=0}^\infty p_e(n, x)s^n \qquad  P_h(s, x) = \sum_{n=0}^\infty p_h(n, x)s^n
\eeq
So multiplying both sides by $s^{n-1}$ and summing from $n=1$ to infinity we find
%
%
%
%
%
%
%
%
%
%
%
%
%
%
%
\beq
      -  \frac{d}{dx} P_e(s, x) = - \alpha P_e(s, x)[1- sP_e(s, x) P_h(s, x)] 
\eeq
\beq
        \frac{d}{dx} P_h(s, x) = -\beta P_h(s, x)[1- sP_e(s, x) P_h(s, x) ]
\eeq
The probability $p(n, x)$ to find $n$ additional e-h pairs for a single e-h pair starting at $x$ is then given by 
\beq
    p(n,x)=\sum_{m=0}^{n} p_e(n-m, x)\,p_h(m, x) \qquad P(s, x) = P_e(s, x) P_h(s, x)
\eeq
With  $P'=P_e'P_h+P_eP_h'$ we therefore find the equation for $P(s, x)$ as
\beq
       \frac{d P(s, x)}{dx } = (\alpha - \beta) P(s, x) [1-sP(s,x)]
\eeq
The equation has the solution
\beq
     P(s, x) = \frac{1}{s\left[1-e^{(\beta-\alpha)x -c_1/s}\right]}
\eeq
and then
\beq
     P_e(s, x) = c_2  [e^{(\beta-\alpha)x}-e^{c_1/s}]^{\frac{\alpha}{\beta-\alpha} }
\eeq
\beq
    P_h(s, x) = c_3  [ e^{(\beta-\alpha)x}-e^{c_1/s} ]^{\frac{-\beta}{\beta - \alpha} }
\eeq
We first note that 
\beq
    P(s, 0) = P_e(s, 0) \qquad  P(s, d) = P_h(s, d)
\eeq
which we use to define $c_2$ and $c_3$ and we have 
\beq
     P_e(s, x) = -(1/s)e^{\alpha x + c_1/s} (1-e^{c_1/s})^\frac{\beta}{\alpha-\beta} (1-e^{(\alpha-\beta)x+c_1/s})^\frac{-\alpha}{\alpha-\beta} 
\eeq
\beq
   P_h(s, x) = -(1/s)e^{-\beta x + c_1/s} 
   (1-e^{(\alpha-\beta)x+c_1/s})^\frac{\beta}{\alpha-\beta} 
   (e^{(\beta-\alpha)d}-e^{c_1/s})^\frac{-\alpha}{\alpha-\beta} 
\eeq
\\ \\
The fact that a single hole at $x{=}0$ and a single electron at $x{=}d$ cannot produce any additional e-h pairs results in the conditions
\beq
    P_h(s, 0) =1 \qquad P_e(s, d) = 1
\eeq
and both of these result in the same equation for $c_1$ 
\beq
       -(1/s)e^{c_1/s} (1-e^{c_1/s} )^\frac{\beta}{\alpha-\beta} (e^{(\beta-\alpha)d}-e^{c_1/s})^{\frac{-\alpha}{\alpha-\beta}} = 1
\eeq
Solving this equation for $c_1$, which will be a function of $s$ i.e. $c_1(s)$ we have our final result for $P(s, x)$. This cannot be done explicitly, so the solution is written in an implicit way. Writing $F(s)=e^{c_1/s}$ we have
%
%
%
\beq \label{elementary_equations}
        P(s, x) = \frac{F(s)}{s(F(s)-e^{(\beta - \alpha)x})} \qquad
         F(s) \left(      \frac{[1-F(s)]^\beta}{[e^{(\beta-\alpha)d}-F(s)]^\alpha}    \right)^\frac{1}{\alpha-\beta} = -s
\eeq
%
%
%
%
%
%
%
%
%
%
%
%
If we are interested in the total number of electron-hole pairs  $\ov p(n, x)=p(n-1, x)$ rather than the number of additional pairs, we have 
\beq \label{PP}
       \ov P(s, x) = s P(s, x ) = \frac{F(s)}{F(s)-e^{(\beta - \alpha)x}}
\eeq
From this point onward, we denote the total-pair distribution
$\overline{p}(n,x)$ again by $p(n,x)$ and its generating
function $\overline{P}(s,x)$ again by $P(s,x)$.
Thus, $n\geq1$ includes the primary electron--hole pair.

\section{Explicit calculation of the probabilities}

\noindent
To find the probabilities $p(n, x)$ we use the Lagrange-B\"urmann inversion formula \cite{Henrici1974} that allows us to find the Taylor expansion of an implicitly defined function.
Assuming a function $F(s)$ that is defined by $ F = s \phi(F)$ with $\phi(0){\neq}0$ and assuming a function $H(F)$, the coefficient of $s^n$ in the Taylor expansion of $H(F(s))$ is equal to $1/n$ times the coefficient of $w^{n-1}$ in the Taylor expansion of $G(w)=H'(w) \phi(w)^n$. \\ 
With the above definitions we have 
\beq
      H(F)=\frac{F}{F-c(x)} \qquad   F=-s\,\phi(F),
  \qquad
  \phi(F)=\frac{(v-F)^\mu}{(1-F)^\nu}.
  \label{eq:F-lagrange-s}
\eeq
\beq
      c(x)=e^{(\beta - \alpha)x} \qquad v=e^{(\beta-\alpha)d}
              \qquad \mu=\alpha/(\alpha-\beta) 
       \qquad \nu = \beta/(\alpha-\beta) 
\eeq
By taking $t=-s$ we have $F=t\phi(F)$ and we can apply the above theorem. With $H'(F)=-c/(c-F)^2$ we find $p(n, x)$ as $(-1)^n/n$ times the coefficient of $w^{n-1}$ in the Taylor expansion of
\begin{equation}
  G(w)=H'(w) \phi(w)^n=
  -\frac{c(v-w)^{\mu n}}{(c-w)^2(1-w)^{\nu n}}
  \label{eq:coefficient-form}
\end{equation}
In the following we use the generalized binomial coefficient
\begin{equation}
\binom{a}{b}
=
\frac{\Gamma(a+1)}
{\Gamma(b+1)\Gamma(a-b+1)},
\end{equation}
Expanding around $w=0$ we have 
\begin{eqnarray}
  (v-w)^{\mu n}
 &=& v^{\mu n}
   \sum_{m=0}^{\infty}
   (-1)^m \binom{\mu n}{m}\frac{w^m}{v^m}.
 \\  (1-w)^{-\nu n}
  &=&\sum_{j=0}^{\infty}
   \binom{\nu n+j-1}{j}w^j.
  \label{eq:second-expansion} \\
  \frac{c}{(c-w)^2}
 & = &
  \sum_{\ell=0}^{\infty}
   \frac{\ell+1}{c^{\ell+1}}w^\ell.
  \label{eq:third-expansion}
\end{eqnarray}
Multiplying these three terms and collecting the coefficients for $w^{n-1}$ we find
\begin{equation} \label{double_sum}
  p(n, x)
  =\frac{(-1)^{n+1}v^{\mu n}}{n}
  \sum_{l=0}^{n-1}\sum_{j=0}^{n-1-l}
  \frac{(-1)^l(n-l-j)}
  {v^l c(x)^{\,n-l-j}}
  \binom{\mu n}{l}
  \binom{\nu n+j-1}{j}.
\end{equation}
This expression can be simplified by transforming the implicit equation for $F$ into an implicit function for the function $U$, which is defined as
\begin{equation}
  F=\frac{vU}{vU-(1-v)}
 \end{equation}
 We then get 
\begin{equation} \label{U_form}
     \ov  H(U)=A\frac{U}{1+\lambda U} \qquad 
      U=t \ov\phi(U)  \qquad
      \ov \phi(U)=(1+U)^\mu 
\end{equation}
 with
\begin{equation}
   t=Cs 
   \qquad
  A(x)=\frac{v}{c(x)(1-v)},
  \qquad
  \lambda(x)=\frac{v(1-c(x))}{c(x)(1-v)}
  \qquad
  C=e^{-\beta d}-e^{-\alpha d}
  \label{eq:A-lambda}
\end{equation}
Again, we find $p(n, x)$ as $1/n$ times the Taylor coefficient of $w^{n-1}$ of 
\beq
  \ov G(w)=\ov H'(w) \ov \phi(w)^n=A
   \frac{(1+w)^{\mu n}}{(1+\lambda w)^2}.
\eeq
and we find
\begin{equation}
  p(n, x)
  =A(x)\frac{C^n}{n}
   \sum_{j=0}^{n-1}
   (-1)^j(j+1)\lambda(x)^j
  \binom{\mu n}{n-j-1}
  \label{second_sol}
\end{equation}
which is now a simpler expression with only a single sum. This expression can be written using the Hypergeometric function ${}_2F_1$ as
\beq
      p(n, x)
  =A(x)\frac{C^n}{n}
   \binom{\mu n}{n-1}
   {}_2F_1\left(
    1-n,2;
    (\mu-1)n+2;
    \lambda (x)
   \right)
     \label{third_sol}
\eeq
We therefore have three equivalent expressions for $p(n, x)$ in Eqs. \ref{double_sum}, \ref{second_sol}, \ref{third_sol}. For $k=1$ the expression becomes independent of $x$ and we have
\beq
     p(n, x) = \frac{(n\alpha d)^{n-1}}{n!}\,e^{-n\alpha d }
\eeq
For $x=0$ we recover the result for 'electron injection only' by McIntyre \cite{McIntyre1972}
\begin{equation}
p(n,0)
=
A(0) C^n\,
\frac{1}{\mu n+1}
\binom{\mu n+1}{n},
\end{equation}
\section{Moments}

\noindent
The $j$th raw moment of the distribution is given by
\begin{equation}
  \langle n^j\rangle
  =\left.
   \left(s\frac{d}{ds}\right)^jP(s,x)
   \right|_{s=1}
  \label{eq:moment-pgf}
\end{equation}
The most efficient way of evaluating this expression is the use of the transformed variable in Eq. \ref{U_form}, where we have 
\beq
   P(s(U),x) = A\frac{U}{1+\lambda U} \qquad   s=\frac{1}{C} \frac{U}{(1+U)^\mu} 
 \eeq
and for $s=1$ we have $U=(1-v)/v$. We can write 
 \beq
     s \frac{d}{ds} = \frac{U(1+U)}{1-(\mu-1)U}\, \frac{d}{dU }
\eeq
and therefore have
\begin{equation}
  \langle n^j\rangle
    =A\left.
   \left[
    \frac{U(1+U)}{1-(\mu-1)U}\frac{d}{dU}
   \right]^j
   \left[\frac{U}{1+\lambda U}\right]
   \right|_{U=(1-v)/v}
\end{equation}
For the first three raw moments we find
\begin{eqnarray}
      \langle n^1\rangle &= & \frac{c(\beta-\alpha) }{\beta-\alpha v}  \\
\left\langle n^2\right\rangle
&=&
\frac{c(\alpha-\beta)^2}
     {(\alpha v-\beta)^3}
\left[
\alpha v(2-v)-\beta
-2(1-c)(\alpha v-\beta)
\right]
\label{eq:second-moment-c}
\\ 
\left\langle n^3\right\rangle 
 &=& \frac{c(\alpha-\beta)}
     {(\alpha v-\beta)^5}
\Bigg[
6c^2(\alpha-\beta)^2(\alpha v-\beta)^2
\\
&+&6c(\alpha-\beta)(\alpha v-\beta)
\left(
\alpha\beta(1-v)^2-(\alpha v-\beta)^2
\right)
\\
&+&(\alpha v-\beta)^4
+2\alpha\beta(\alpha+\beta)(1-v)^3
 (\alpha v-\beta)
+3\alpha^2\beta^2(1-v)^4
\Bigg]
\label{eq:third-raw-moment-c-alpha-beta-v}
\end{eqnarray}

\section{Mean, variance, third central moment}

\noindent
For the discussion we use the traditional replacement of $\beta=k\alpha$. The first moment represents the average gain, which we write as $M(x)=  \langle n^1\rangle$
\begin{eqnarray}
  M(x) = M(0) e^{-(1-k)\alpha x} = M(d) e^{(1-k)\alpha (d-x)}
\end{eqnarray}
where $M(0)$ and $M(d)$ are the gains for electron injection at $x{=}0$ or hole injection at $x{=}d$. 
The condition for finite gain is $d{<}\ln (\alpha/\beta)/(\alpha-\beta)$, so only under this condition the probability distributions are normalized. For $k{=}1$ this condition is $\alpha d {<}1$.
Expressing $v$ by $M(0)$ or $M(d)$ we can write the second raw moment as
\begin{eqnarray}
\left\langle n^2\right\rangle
& = &
2M(x)^2
+
M(x)
\left[
k(M(0)-1)^2-1
\right] \\
& = &
2M(x)^2
+
M(x)\left[
\frac{(M(d)-1)^2}{k}-1
\right]
\label{eq:second-moment-Mx-M0}
\end{eqnarray}
The expressions are equal to the ones from \cite{Zambon2024} and for $x{=}0$ and $x{=}d$ they recover the ones from McIntyre \cite{McIntyre1966}. The variance is then $ \sigma(x)^2=\left\langle n^2\right\rangle - M(x)^2$.
%
%
%
%
%
%
%
%
%
%
%
%
%
%
%
%
%
%
%
%
%
The third central moment $T(x)$ is  defined as 
\begin{equation}
T(x)
=
\left\langle
\bigl(n-M(x)\bigr)^3 \right\rangle=\left\langle n^3\right\rangle
-3M(x)\left\langle n^2\right\rangle
+2M(x)^3.
\label{eq:third-central-moment-definition}
\end{equation}
Inserting the above expressions we find
\begin{eqnarray}
T(x)   \nonumber
&=&
2M(x)^3
+3\left[k(M(0)-1)^2-1\right]M(x)^2
+
\left[
1+2k(1+k)(M(0)-1)^3
+3k^2(M(0)-1)^4
\right]M(x) \\ \nonumber
&=& 2M(x)^3
+3\left[
\frac{(M(d)-1)^2}{k}-1
\right]M(x)^2
+
\left[
1+
\frac{
2(1+k)(M(d)-1)^3
+3(M(d)-1)^4
}{k^2}
\right]M(x).
\label{eq:third-central-moment-M-M0}
\end{eqnarray}
which is equivalent to the expressions derived in \cite{Zambon2024}.

\section{Solution for $m$ primary electron-hole pairs at position $x$}

\noindent
If instead of a single e-h pair we have $m$ primary e-h pairs at position $x$, the generating function $P_m(s, x)$ for the probability $p_m(n,x)$ to find $n$ total e-h pairs (including the $m$ primary ones) is 
\beq
P_m(s, x) = P(s,x)^m
=
A^m\frac{U^m}{(1+\lambda U)^m}.
\eeq
Proceeding as before with the Taylor expansion we get the two equivalent expressions for $p_m(n, x)$
\begin{eqnarray} \nonumber
p_m(n, x)
&=&{}
\frac{mA(x)^mC^n}{n}
\sum_{j=0}^{n-m}
(-1)^j
\lambda(x)^j
\binom{m+j}{j}
\binom{\mu n}{n-m-j} .
\\
&=&{}
\frac{mA(x)^mC^n}{n}
\binom{\mu n}{n-m}
\times
{}_2F_1\left(
m-n,\,
m+1;\,
(\mu-1)n+m+1;\,
\lambda (x)
\right).
\label{final}
\end{eqnarray}
with
\beq
     p_m(n, x) = 0 \qquad \mbox{for} \qquad n<m
\eeq
For $x{=}0$ we again recover the expression from McIntyre \cite{McIntyre1972}.
For $k{=}1$ the expression becomes 
\beq
     p_m(n, x) = \frac{m}{n} \frac{(n\alpha d)^{n-m}}{(n-m)!}\,e^{-n\alpha d }
\eeq
For the mean, the variance and the third central moment we have
\beq
     M_m(x) = m \, M(x) \qquad  \sigma_m(x)^2 = m \, \sigma(x)^2 \qquad T_m(x)= m \, T(x)
\eeq

\section{Example}

\begin{figure}[ht]
 \begin{center}
  \includegraphics[height=10cm]{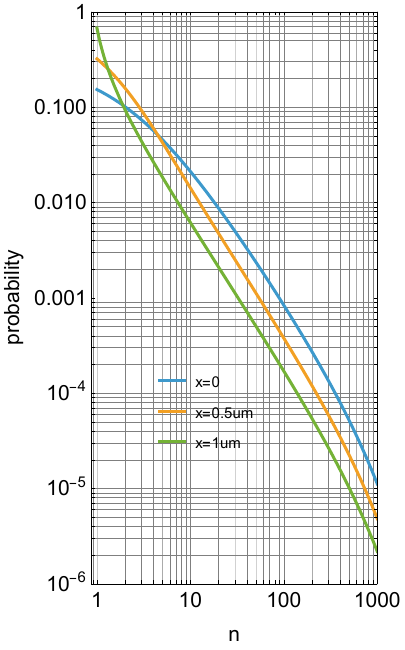}
  \includegraphics[height=10cm]{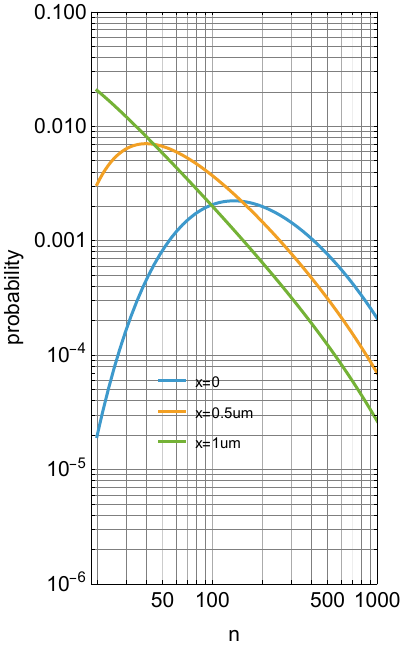}
  \caption{Probability of finding $n$ e-h pairs for 1 primary e-h pair at $x{=}0,0.5,1\,\mu$m (left) and 10 primary e-h pairs at $x{=}0,0.5,1\,\mu$m (right). }
  \label{avalanche_distribution}
  \end{center}
\end{figure}

\noindent
As an example, consider diode of $1\,\mu$m thickness with an applied voltage of 34\,V.
Using the model of \cite{VanOverstraeten1970} we have $\alpha {\approx} 1.88 \mu$m$^{-1}$ and $\beta  {\approx}  0.4 \mu$m$^{-1}$, so $k{\approx} 0.21$ and the gain $M(0) {\approx} 50$ for an e-h pair at $x{=}0$ and $M(d){\approx}11$ for an e-h pair at $x{=}d$. The total avalanche size distributions for 1 and 10 primary e-h pairs at $x{=}0, 0.5, 1\,\mu$m are given in Fig. \ref{avalanche_distribution}.

\clearpage

\section{Position dependence of $\alpha$ and $\beta$}

\noindent
IIn a nonuniform diode, the ionization coefficients
$\alpha(x)$ and $\beta(x)$ depend on position, and a simple
closed-form expression for the complete avalanche-size
distribution is generally not available. Its moments can
nevertheless be calculated iteratively, as shown, for example,
in Ref.~\cite{Zambon2024}. If the ionization coefficients
satisfy
$
\beta(x)=k\alpha(x),
$
with a position-independent constant $k$, all the results
derived above can be generalized in a straightforward manner. The solution of 
\beq
       \frac{d P(s, x)}{dx } = \alpha(x)(1-k) P(s, x) [1-P(s,x)]
\eeq
is 
\beq
     P(s, x) = \frac{1}{1-e^{-(1-k)\int_0^x \alpha (u) du -c_1/s}}
\eeq
and from there, the calculations proceed in the same way as the ones for constant $\alpha$ and $\beta$. As a result we just have to make the following replacements for the expressions for constant $\alpha$ and $\beta$:
\beq
     \alpha x \,\rightarrow \,a(x)= \int_0^x \alpha(u)du \qquad \alpha d \, \rightarrow \, a(d)
\eeq
Specifically we then have
\beq
   c(x)=e^{-(1-k)a(x)} \qquad
   v=e^{-(1-k)a(d)} \qquad 
    \mu = \frac{1}{1-k} \qquad 
     \nu = \frac{k}{1-k}  \qquad
\eeq
\beq
       C=e^{-ka(d)}-e^{-a(d)} \qquad
A(x)=\frac{v}{c(x)(1-v)},
\qquad
\lambda(x)=\frac{v[1-c(x)]}{c(x)(1-v)}.
\eeq
Since the proportionality condition $\beta(x)=k\alpha(x)$ is generally not well satisfied in silicon, the applicability of this more general result is limited.

\section{Conclusion}

\noindent
The key result of this report is the probability distribution for the avalanche size in a uniform avalanche diode of thickness $d$ for $m$ primary e-h pairs at an arbitrary position $0\le x \le d$ inside the sensor (Eq. \ref{final} ). For $x{=}0$ and $x{=}d$ the expression recovers the results from McIntyre \cite{McIntyre1972}. 

\clearpage




\end{document}